# Title: An experimental demonstration of neuromorphic sensing of chemical species using electro-optical reservoir computing

*Gleb Anufriev[1], David Furniss[1], Mark C. Farries[1,2], Angela B. Seddon[1] and *Sendy Phang[1]

[1] Mid-infrared Photonics Group: George Green Institute for Electromagnetics Research Faculty of Engineering, University of Nottingham, Nottingham, NG7 2RD, UK

[2] Marine Photonics LTD. Deepdene Cottage, Deepdene Park, Exeter, England, EX2 4PH

*Contact Authors: gleb.anufriev@nottingham.ac.uk   sendy.phang@nottingham.ac.uk

**Abstract**

A chemical discrimination system based on photonic reservoir computing is demonstrated experimentally for the first time. The system is inspired by the way humans perceive and process visual sensory information. The electro-optical reservoir computing system is a photonic analogue of the human nervous system with the read-out layer acting as the 'brain', and the sensor that of the human eye. A task-specific optimisation of the system is implemented, and the performance of the system for the discrimination between three chemicals is presented. The results are compared to the previously published numerical simulation [1]. This publication provides a feasibility assessment and a demonstration of a practical realisation of photonic reservoir computing for a new neuromorphic sensing system - the next generation sensor with a built-in 'intelligence' which can be trained to 'understand' and to make a real time sensing decision based on the training data.

## 1. Introduction

The human nervous system possesses remarkable computational abilities [2,3]. It is an incredibly powerful biological computer capable of performing pattern recognition, regression and forecasting on massively parallel information in real time [4,5]. Inspired by this, artificial neural networks (ANNs) aim to replicate the neural functions of humans as a computational framework. Such systems are called 'neuromorphic', as they are inspired by the computing architecture of the human nervous system and brain. ANNs and neuromorphic systems have shown effectiveness for tasks, like pattern recognition and inference, and found applications in bioinformatics, medical image processing, stock market forecasting, and telecom signal recognition [6-16].

Numerous architectures, implementations, and applications of ANNs have been demonstrated – both in software and as hardware systems [17-19]. Among those, photonic implementations have been demonstrated to be suitable for high-speed processing due to the larger bandwidths offered by optical signals and components [19,20]. Photonic reservoir computing (PhRC) is one such implementation. This offers an alternative approach to the architecture and functionality of photonic neural networks [21-23], which are based on the conventional feed-forward neural networks. In reservoir computing, which was first demonstrated as a software implementation, training is exclusively carried out at the read-out layer, allowing the kernel to remain semi-random and untrained [19, 21-23]. The reservoir kernel (see Fig. 1(b)) performs complex temporal dynamics and nonlinear transformations which processes input data, remapping it to a new higher-dimensional representation space [24-26]. This higher-dimension representations allows for a final linear discrimination to be performed by the read-out layer. PhRC systems have been demonstrated in an optical-fibre setup, including electro-optical feedback [25,26], all-optical feedback [27,28] and all-optical stimulated Brillouin scattering systems [29], in a chip-scale integrated photonic devices, including as network of complex interconnected waveguides [30], stochastic photonic field [31,32], and lasers [18,33]. Using these PhRC platforms, applications for telecommunications, quantum computing, and chaotic time series generation and prediction have already been reported [21,22].

In the present work, we considered the practical implementation of electro-optical reservoir computing (EORC), with an optical fibre delay-line and Mach-Zehnder modulator, which provide the memory and nonlinear effects respectively [1,25,26]. The work presented here focuses on the first experimental demonstration of PhRC as neuromorphic chemical sensing system - the next generation of sensing with built-in intelligence. Such a system can be trained to make real time sensing decisions based on training data - inspired by the fact that the human computing capacity is predominantly used for processing of sensory information [34]. The sensory part of the neuromorphic sensing system was implemented in the infrared, mimicking the properties of human eyes, namely the discrete and broadband response of the cone cells in the retina. The application of this neuromorphic sensory system was demonstrated to the discrimination of different chemicals.

This paper is structured as follows: Section 2 describes the neuromorphic sensory system. This starts by outlining the experimental operation of the pyroelectric sensing apparatus, describing the experimental implementation of the EORC, and the pre- and post-processing of data that was carried out. Section 3 presents and discusses the results obtained and their comparison to the earlier work [1]. Section 3, further, reports the dynamics of the EORC kernel, before presenting the results on stability and performance for the chemical discrimination tasks. The impact of the optimisation parameters on system performance is also briefly discussed. The conclusions are provided in Section 4.

## 2. Methods

In this Section, the experimental implementation of the neuromorphic sensing system is described. Figure 1(a) describes the schematic of operation of the sensing system. The neuromorphic sensing system comprises four distinct layers (see Fig. 1(c)) – the sensing layer, the EORC kernel, the read-out layer and the control layer. Sections 2.1 and 2.2 describe the sensing layer which consists of a pyroelectric sensor, and data pre-processing and generation. Figure 1(b) shows the reservoir computing kernel and the algorithm used for chemical classification. Section 2.3 describes the physical implementation and components of the EORC shown in Fig. 1(c). This Section also describes the control layer – which was used to optimise the EORC through particle swarm optimisation. The state of the signal at every stage of the EORC is not described in detail here, and readers are referred to the simulation counterpart of the paper [1], but the data post-processing which happens at the read-out layer, is described in Section 2.4.

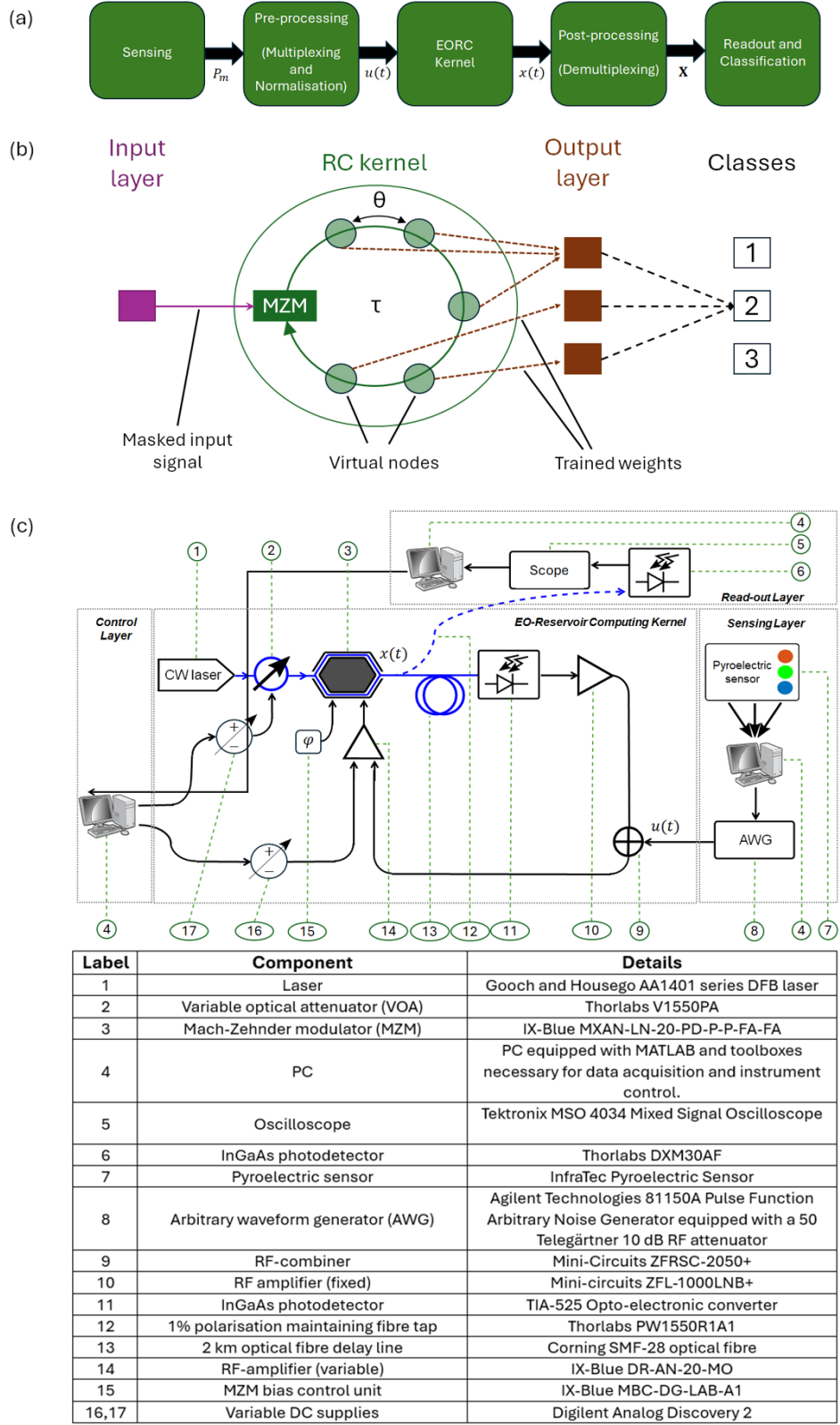

*Figure 1: (a) Flow diagram for the process of chemical discrimination by the neuromorphic sensing system described in this Section. (b) Delayed feedback reservoir computing kernel architecture and the classification algorithm. The experimental parameters τ and θ are the feedback-delay time and duration for each mask value respectively (See: Section 2.2). (c) The experimental setup developed in the current work implementing the delayed feedback reservoir computing and a summary of components involved in the experimental setup. Variable DC*

*sources were used to tune the attenuation of the variable optical attenuator and the gain of the RF-amplifier driven by particle swarm optimisation.*

2.1 The pyroelectric sensor

A pyroelectric sensor equipped with a tuneable broadband Fabry-Perot filter [35] was used for recording the transmissive spectral response of chemical samples, see Fig. 2(a). This sensor produced discrete and broadband spectral responses. The central wavelength of the Fabry-Perot filter, $\lambda_c$, was tuneable over a wavelength range of 3000 nm $\leq \lambda_c \leq$ 4500 nm with a full-width half-maximum (FWHM) of 47.65 nm as depicted in Fig. 2(b). The sensor used a broadband thermal blackbody source. The spectral response of the sensor was obtained by sweeping the central wavelengths, $\lambda_c$, of the Fabry-Perot filter across the spectral region of interest, from which a low-resolution transmittance spectrum for each sample was obtained. The normalised transmittance, $T$, describes a normalised detected power, $P_m$, obtained from the sensor with respect to the reference background with no-sample present. This can be expressed mathematically as,

$$T(\lambda_c) = \frac{\int S_{\text{in}}(\lambda) D(\lambda, \lambda_c)(1 - A(\lambda)) d\lambda}{\int S_{\text{in}}(\lambda) D(\lambda, \lambda_c) d\lambda} \tag{1}$$

where, $S_{\text{in}}(\lambda)$ is the spectral power density of the source, $D(\lambda, \lambda_c)$ is the spectral response of the Fabry-Perot filter tuned at centre wavelength, $\lambda_c$, and $A(\lambda)$ denotes the spectroscopic absorption of the sample-under-test. It is important to point out, as can be deduced from Eq. (1), that the transmittance of the pyroelectric sensor, $T$, was not equal to the spectroscopic transmittance, i.e., $T(\lambda_c) \neq 1 - A(\lambda)$.

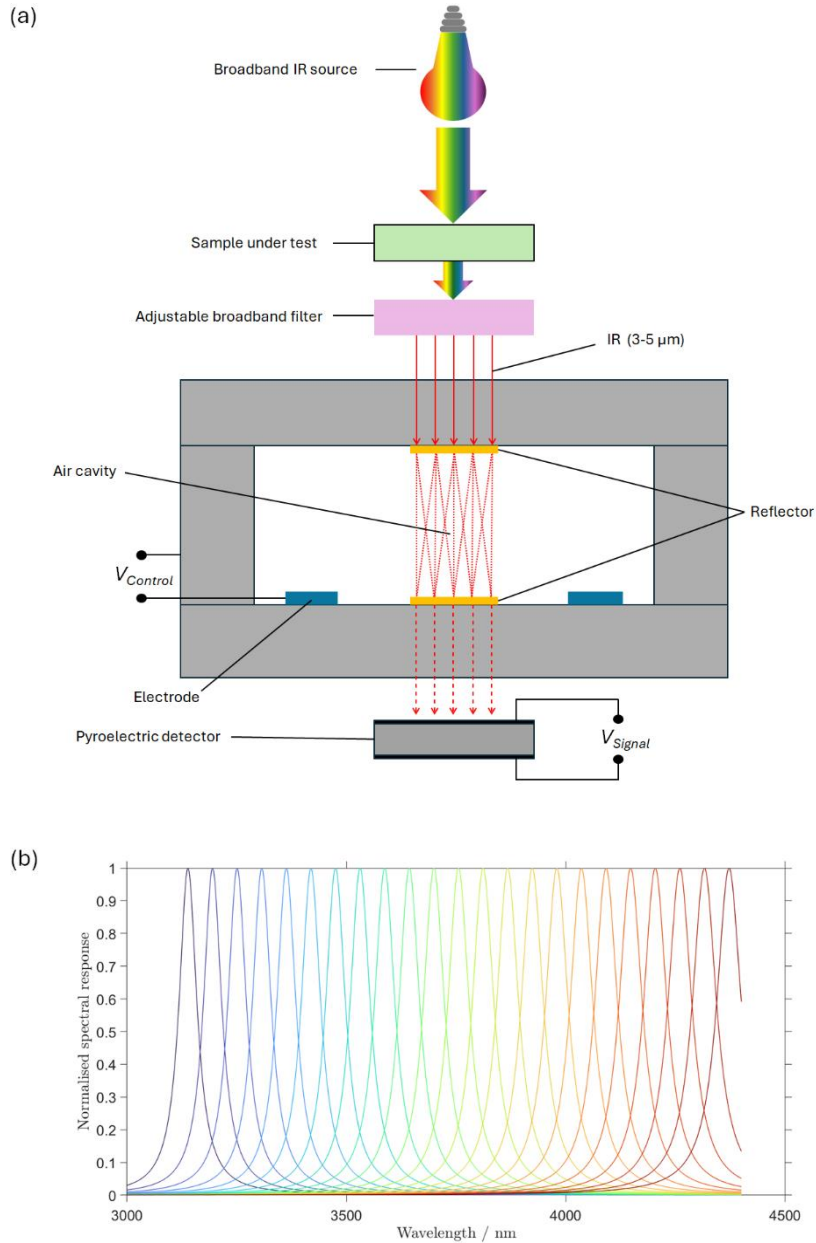

*Figure 2 (a) Schematic of the pyroelectric sensor. (b) The Fabry-Perot filter spectral response.*

Motivated by human visual perception, which is capable of discriminating over 100,000 shades of colours from only three colour sensitive retinal neuron ends (trichromat arrangement [36,37]), in this present work only three central wavelengths, $m$, of the Fabry-Perot filter, i.e. $\lambda_{c,m}$ where $m = 1,2,3,$ were considered for each chemical sample. A set of three selected central wavelengths, $\lambda_{c,m}$, was chosen for each group of the samples-under-test, aiming at maximising the difference between $T(\lambda_{c,m})$ within the group. The transmittance obtained from the sensor, as well as the three central wavelengths, $\lambda_{c,m}$, selected for the chemical discrimination of each group, are depicted in Fig. 3(a) for the group of aliphatic alcohols and Fig. 3(b) for the group of essential oils. The central wavelengths, $\lambda_c$, selected are marked with dashed lines. For the group of aliphatic alcohols (Fig. 3(a)) the selected central wavelengths were 3140 nm, 3280 nm and 3460 nm. For the group of essential oils (Fig. 3(b)) the selected central wavelengths were 3130 nm, 3430 nm and 3700 nm. The spectra depicted in Fig. 3 were collected at different times over a period of a week, confirming robustness and reproducibility of the pyroelectric sensing apparatus data, and the standard deviation of the transmittance spectral data, $T(\lambda_{c,m})$, is shown as error-bars. This transmittance dataset was digitally pre-processed and normalised, to be used as input signal to the EORC kernel.

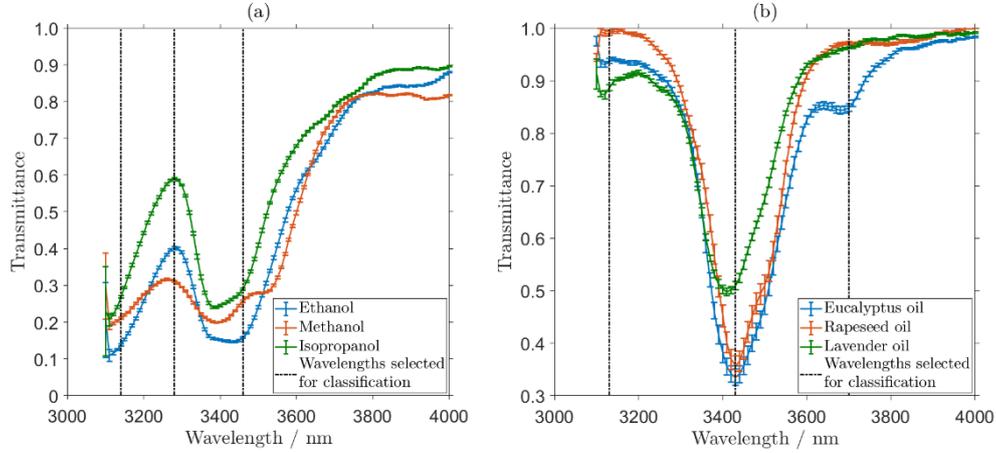

*Figure 3: The transmittance dataset obtained using the pyroelectric sensor for (a) the group of aliphatic alcohols and (b) the group of essential oils. The error-bars represent the standard deviation of the obtained responses, $T(\lambda_{c,m})$. The wavelengths that were selected for the discrimination algorithm are depicted by black dashed lines.*

2.2 Pre-processing: information representation and time-multiplexing

In order for the discrete transmittance data from the sensor to be used, it was represented in a format suitable for the EORC kernel through data pre-processing. The first step in pre-processing the raw spectral transmittance data was a sample-and-hold of this dataset, only considering the transmittance at three distinct central wavelengths (see Fig. 3) $\boldsymbol{\mathcal{T}} = [T(\lambda_{c,1}), T(\lambda_{c,2}), T(\lambda_{c,3})]$, producing an analogue RF signal, $j(t)$. Figure 4(a) illustrates the sample-and-hold procedure in which each transmittance data value, $T(\lambda_{c,m})$, was held for a duration $\tau$, the feedback round-trip delay of the EORC. Thus, resulting from the three central wavelengths considered in this work, $j(t)$ was a generated radio frequency (RF) information signal with period of $3\tau$. To exemplify this, the information signal, $j(t)$, for a single spectrum of ethanol is given in Fig. 4(a).

The subsequent step in data pre-processing was time-multiplexing. Time-multiplexing is a common approach used in EORC to increase signal diversity, allowing for an increased number of virtual neural nodes from a single physical node [25,26]. Time-multiplexing was achieved through the imposition of a periodic mask, $m(t)$, to the serialised information signal $j(t)$ [1]. Figure 4(b) shows an example of mask signal, $m(t)$; The mask signal was a flat top function with a periodicity of $\tau$, constructed from a series of random numbers. These random numbers were generated by a random number generator with a normal distribution, a mean of 1 and a scaling factor of 0.3, i.e., $m(t) \in [0.85; 1.15]$ with $\tau = N_x \theta$, where $N_x$ is the number of mask values and $\theta$ is the duration for each mask value. The scaling factor applied to the random number generator was experimental consideration which allowed all individual mask values to be resolved within the capability range of the arbitrary waveform generator and oscilloscope.

The input signal, $u(t)$, to the EORC, derived from the pre-processed raw spectral acquisition, was then achieved through the modulation of the serialised information signal, $j(t)$, by the mask signal, $m(t)$. An arbitrary waveform generator (AWG), labelled (8) in Fig. 1(c), was used to produce and feed the input signal, $u(t)$, to the EORC kernel. Figure 4(c) shows an example input signal, $u(t)$, for a single transmittance measurement of ethanol with $N_x$ = 5.

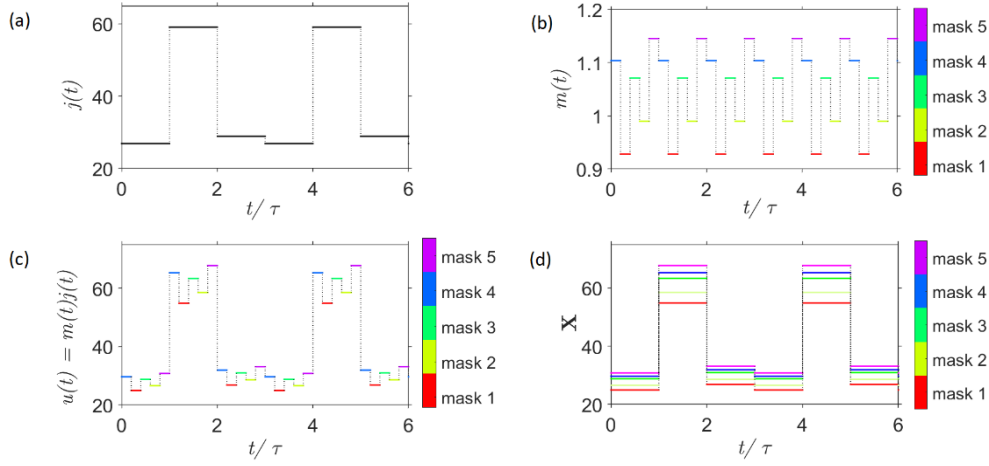

*Figure 4. (a) An example of a serialised information signal, $j(t)$, obtained for a single instance of transmittance measurement of ethanol $\mathcal{T} = [T(\lambda_{c,1}), T(\lambda_{c,2}), T(\lambda_{c,3})]$. The signal is periodic with $3\tau$, where $\tau$ is the round-trip time for the EORC. (b) A depiction of a mask signal, $m(t)$, with 30% modulation depth and a bias of 1. The mask signal is periodic with $\tau$ and each mask lasts for $\theta = \tau/N_x$, where $N_x$ is the number of random value masks (5 in this example). (c) An example of the EORC input signal, $u(t)$, created from the modulation of the serialised information signal, $j(t)$, by the mask signal, $m(t)$. The depicted example is for a single instance of an ethanol transmittance measurement with 5 mask values, shown in different colours. (d) A visualisation of demultiplexing for an input signal, $u(t)$, with $N_x = 5$. Application of demultiplexing to the reservoir activation states, $x(t)$, instead, yields the neuron activation states (**X**).*

2.3 Electro-optical reservoir computing (EORC) system

The EORC implementation used a continuous-wave laser operated at 1560 nm, with a maximum power output of 100 mW, labelled (1) in Fig. 1(c). The laser provided a carrier-signal which propagated through a 2 km single-mode, SMF-28, fibre feedback delay-line system. The laser signal was modulated by a combination of an analogue input signal, $u(t)$, and the feedback signal propagating through the system. This was referred to the synchronous regime of operation in previous work [26]. These signals were combined using an RF-combiner labelled (9) in Fig. 1(c), and fed through a variable RF-amplifier, labelled (14) in Fig. 1(c), before being used as an RF-modulation input to the Mach-Zehnder modulator (MZM). In this way, the EORC kernel had two physical optimisation parameters, i.e. the laser power $P$ and the signal amplification $\gamma$. Suitable kernel parameters were obtained through a particle swarm optimiser, implemented within MATLAB [38]. The laser power was controlled through a DC signal, supplied to an electronic variable optical attenuator, labelled (2) in Fig. 1(c). To compensate for component temperature drifts and obtain a consistently precise and accurate carrier signal power, a simple proportional-integral-derivative (PID) controller was implemented. Similarly, the signal amplification was also controlled through the supply of a DC signal to a RF-amplifier, labelled (14) in Fig. 1(c). For detailed information on particle swarm optimisation, readers are referred to the particle swarm optimisation toolbox [38].

Although an additional optimisation parameter could also be found from the MZM operation, in the present work, the MZM was operated at quadrature, as previous works suggested that optimal performance of the EORC signal discrimination task was obtained in this regime of operation [1,26]. The negative quadrature point was selected, as it was observed that our MZM device exhibited robust operation for DC-bias voltages closest to 0 V. Negative quadrature was obtained at a DC-bias of -2 V, while positive quadrature was observed at a DC-bias of +4 V. The round-trip time for the EORC circuit, $\tau$, was measured as 10.98 μs by recording the pulse delay at the read-out layer. Readers are referred to the systematic analysis of the EORC system in previous publications [1,25,26].

2.4 Post-processing: time-demultiplexing, training and testing at the read-out layer

The signal from the EORC kernel, i.e., the reservoir activation states, was monitored using at an InGaAs photodetector, labelled (6) in Fig. 1(c), obtained at the read-out layer feeding into an oscilloscope, labelled (5) in Fig. 1(c), at a 1% power optical tap, labelled (12) in Fig. 1(c). An oscilloscope was used in our setup to record an averaged steady state of the system and send this to a computer for data post-processing, see Fig. 1(c).

The first step in data post-processing was demultiplexing. Its purpose was to deconvolute the EORC activation states at the read-out layer, and thus, obtain the individual neuron activation states based on the virtual RC nodes approach. This procedure is most simply shown when considering a system with no feedback and a small input signal, $u(t)$; however, the procedure is the same for a system with feedback and any $u(t)$ (see Fig. 4(d)). For a system with no feedback and a small input signal, $u(t)$, the reservoir activation states, $x(t)$, observed at the read-out were the same as $u(t)$ (see Fig. 4(c)). Since $u(t)$ was created from a periodic perturbation of the serialised information signal, $j(t)$, by the mask, $N_x$ neuron activation states, **X**, were obtained, where $N_x$ is the number of masks. The resulting activation states **X** were then used in the subsequent training and testing steps. In the case of a system with a small input signal and no feedback, these activation states all resembled $j(t)$. In systems with feedback, the neuron activation states, **X**, do not resemble the serialised information signal. An example of the demultiplexing process can be seen in Fig. 4(d) for a system with a small input signal and no feedback.

The final step in the data post-processing depended on the regime of operation of the neuromorphic sensing system, i.e. training or testing. During training, the read-out of the EORC system 'learned' about the task by evaluating the output weight, $\mathbf{W}_\text{out}$, such that the difference between the known $\mathbf{Y}_\text{training}$ and the matrix of neuron activation states, **X**, was minimised. This error minimisation-based learning was achieved by Tikhonov regularisation with cross-validation as: [39]

$$\mathbf{W}_\text{out} \leftarrow \min[\|\mathbf{W}_\text{out}\mathbf{X}^\text{T} - \mathbf{Y}_\text{training}\|] \tag{2}$$

where the desired output, **Y**, is a numerical representation of the sample-under-test using one-hot-encoding, $\mathbf{Y}_\text{training}$ refers to the desired outputs used during the training and $(\cdot)^\text{T}$ denotes the transpose operator. One-hot-encoding used to represent target chemicals is further detailed in Section 3.2.

The neuromorphic sensing system was operated in the testing regime after being trained. In the testing regime, the read-out layer produced prediction signals, $\mathbf{Y}_{\text{prediction}}$, from the $\mathbf{W}_{\text{out}}$ obtained while training, and a set of neuron activation states, $\mathbf{X}$, for the sample-under-test. Mathematically, this can be expressed as in Eq. (3). It should be noted that the transmittance datasets were split for each chemical such that the system was trained and tested on different transmittance data.

$$\mathbf{Y}_{\text{prediction}} = \mathbf{W}_{\text{out}} \mathbf{X}^{\text{T}} \tag{3}$$

The overall system performance, i.e., the error rate of the prediction, was evaluated by comparing the predicted signal outputs, $\mathbf{Y}_{\text{prediction}}$, to the target outputs, $\mathbf{Y}_{\text{target}}$. In this work, the normalised mean squared error ($NMSE$) defined as the following was used throughout as a metric to evaluate performance,

$$NMSE = \frac{\langle \|\mathbf{Y}_{\text{prediction}} - \mathbf{Y}_{\text{target}}\|^2 \rangle}{\text{var}(\mathbf{Y}_{\text{target}})}, \tag{4}$$

where $\text{var}(\cdot)$, $\langle \cdot \rangle$ and $\|\cdot\|$ denote the variance, the assemble averaging, and the Euclidean norm operators, respectively.

## 3. Results and Discussion

This Section starts with the characterisation of the EORC with no sensory input present. Section 3.2 demonstrates the application of the neuromorphic sensing system to the discrimination of chemicals, describes the performance optimisation of the EORC, and the system stability.

3.1 EORC operation states and bifurcation.

The EORC used in the present work was based on a delay feedback system [1,25,26]. Such a system features parameter dependent behaviour and was characterised by the bifurcation diagram.

System bifurcation was achieved in our setup by varying input laser power ($P$) and the gain of the RF-amplifier ($\gamma$), labelled (1) and (14) in Fig. 1(c) respectively. First, consider Fig. 5(b), which depicts the bifurcation of the monitored optical signal at the read-out photodetector (labelled (6) in Fig. 1(c)). To illustrate this bifurcation phenomenon, the states of the optical signal, i.e., the $\max[x(t)] - \min[x(t)]$, have been plotted. For a fixed input laser power of $P_0$ = 55 mW, and low amplification, the states of the optical signal remained constant (single-valued), i.e., $\max[x(t)] \approx \min[x(t)]$, as the gain of the RF-amplifier, $\gamma$, increased. This is because, although the gain $\gamma$ increased, the optical power provided by the laser module remained constant. In the single-valued region, the small fluctuation observed was caused by random noise, which was observed to be around 0.2 V. However, after a specific RF-amplifier gain of $\gamma = 23.4$ dB, the states of the optical signal distinctly split denoting the bifurcation. Figure 5(c) shows the bifurcation of the optical signal as a function of the input laser power ($P$) for a fixed RF-amplifier gain $\gamma$ = 25 dB. Figure 5(c) demonstrates a similar bifurcation behaviour, but with an overall positive gradient due to the increase of overall input laser power. To further illustrate the bifurcation phenomenon, examples of the temporal signals of the system are shown in Figs. 5(d), (e) and (f) for the specific input laser powers ($P$) marked in Fig. 5(c).

To gain an overall picture of the bifurcation phenomenon, the surface plot in Fig. 5(a) shows the state of the optical signal as a function of both the input laser power ($P$) and the RF-amplifier gain ($\gamma$). Figure 5(a) shows two distinct regions: the region where the system response was single-valued (dark blue) and the region where the system response was oscillating (not-blue). This bifurcation behaviour, single-valued for both low input laser power ($P$) and the RF-amplifier gain ($\gamma$); and oscillating for high input laser power ($P$) and the RF-amplifier gain ($\gamma$), is consistent with previously published simulation work [1]. The transition between the two states, the bifurcation line, is indicated by the light blue boundary between dark blue and green colours in Fig. 5(a). Furthermore, the specific parameters used for Figs. 5(b, c), namely $P_0$ = 55 mW and $\gamma$ = 25 dB, are marked in Fig. 5(a) by dashed white lines. The method used in characterising the different regions of operation was a two-parameter sweep across laser power ($P$) and RF-amplifier gain ($\gamma$). This only allowed the location of the first order bifurcation points (the first instance of the system response becoming multi-valued). It is possible that higher order bifurcations exist, but they were not considered in this paper.

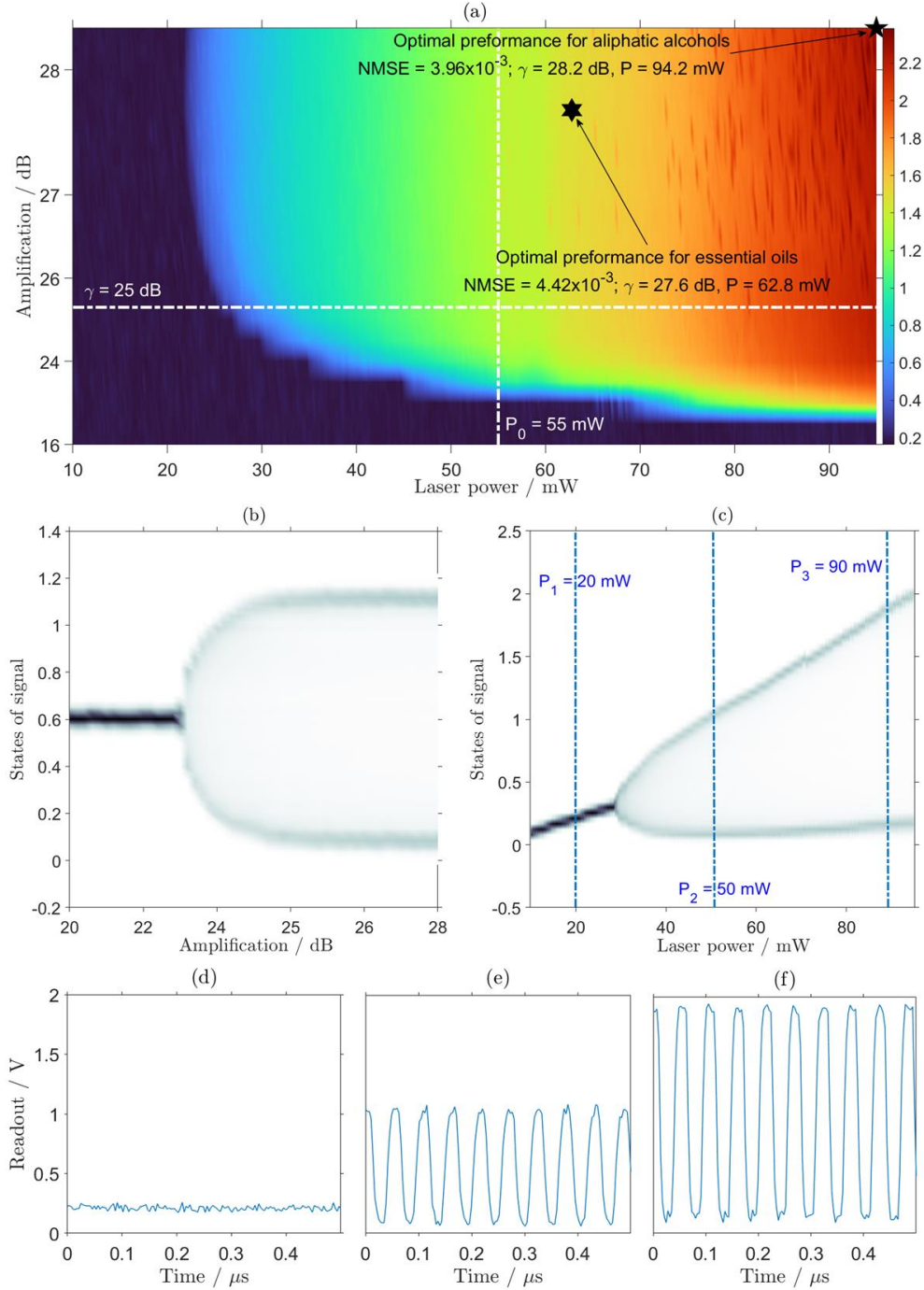

*Figure 5. Operational states of EORC kernel. (a) The range of reservoir activation states, $x(t)$, observed at the read-out photodetector as function of EORC parameters: laser power P and RF-amplifier gain $\gamma$. The single-valued system response is shown in dark blue and the region where the system response was oscillating is shown with other colours. The colour bar depicts the difference in Volts between the maximum value of x(t) and the minimum value of x(t) (i.e. the states of the signal) observed on the oscilloscope. The dashed lines mark the specific input laser power $P_0 = 55$ mW and RF-amplifier gain $\gamma = 25$ dB, for the bifurcation diagram (b) and (c), respectively. The star marks the optimum operation region obtained using particle swarm optimisation for the discrimination of a group of aliphatic alcohols. The hexagon marks the optimum operation region obtained using particle swarm optimisation for the discrimination of a group of essential oils (see Section 3.2). (d), (e), and (f) depict the temporal signals $x(t)$ for the*

*specific operation parameters at amplifier gain $\gamma = 25$ dB , $P = 20$ mW, $50$ mW, and $90$ mW as marked in (c).*

### 3.2 Chemical discrimination by the neuromorphic sensing system

A group of three aliphatic alcohols, i.e. ethanol, methanol, and isopropanol, was first used for chemical discrimination to demonstrate the accuracy of the trained system, allowing for direct comparison with previously published simulation work [1]. Furthermore, to show its universal application of handling other chemical samples, the sensing system was then trained to classify a group of essential oils - eucalyptus, lavender, and rapeseed oils. A dataset of 90 spectral responses was used during the discrimination of the group of aliphatic alcohols (30 spectra for each type of alcohol), and 60 spectra during the discrimination of the group of essential oils (20 spectra for each type of oil). In both cases, 80% of the spectra available were used for training and 20% for testing.

As described in Section 2.4, one-hot encoding was used to represent the chemical sample-under-test, here, it is defined as in Table 1.

Table 1: One-hot-encoding for discrimination of groups of chemicals.

| Discrimination of: | Aliphatic alcohols | Essential oils |
|---|---|---|
| $\mathbf{Y}_{\text{target}} = \begin{cases} [1,0,0] \\ [0,1,0] \\ [0,0,1] \end{cases}$ | Ethanol<br>Methanol<br>Isopropanol | Eucalyptus oil<br>Rapeseed oil<br>Lavender oil |

Particle swarm optimisation is a very common optimisation used in engineering [40,41]. The details of the particle swarm optimisation approach can be found in the documentation for the particle optimisation toolbox [38]. Heuristic optimisation using the particle swarm method was employed to find the optimum EORC operational parameters for accurate sample classification. This approach yielded a minimised the *NMSE* between the target ($\mathbf{Y}_{\text{target}}$) and the predicted ($\mathbf{Y}_{\text{prediction}}$) signals. The particle swarm optimisation for this minimisation was set to use 200 particles at each iteration. At each iteration, the set of EORC operation parameters, $P$ and $\gamma$, which yielded the lowest value of *NMSE* was recorded. For the aliphatic alcohol discrimination, the optimisation concluded after 50 iterations (Fig. 6(a)) and after 20 iterations for the discrimination of essential oils (Fig. 6(b)). The EORC operation parameters, $P$ and $\gamma$, that corresponded to the lowest recorded *NMSE* are marked in Fig. 5(a) by a star for discrimination within a group of aliphatic alcohols and a hexagon for the discrimination within a group of essential oils. The *NMSE* for classification of chemicals in each group at optimal parameters is shown in Fig. 5(a), $N_x$ = 50 mask values were used for this task. It is noted that the optimum operation point for aliphatic alcohol discrimination was not in line with the previously carried out simulation [1], where it was shown to be near the bifurcation points. We believe that this is due to our use here of the particle swarming method, which is based on stochastic optimisation, converging to a local minimum, in contrast to the exhaustive search conducted in the simulation work [1]. This suggests that further system performance improvements could be possible.

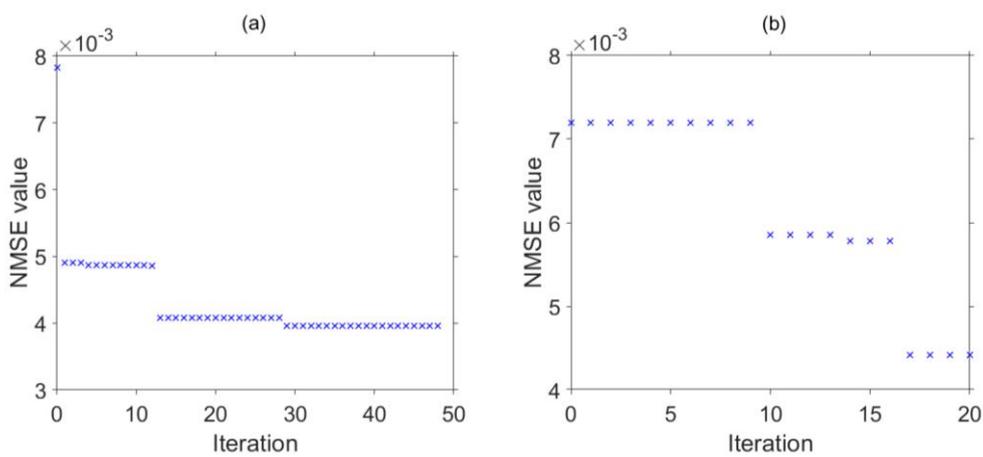

*Figure 6. EORC operational parameter optimisation by the particle swarm method. Averaged classification NMSE with each particle swarm optimisation iteration for: (a) the group of aliphatic alcohols and (b) the group of essential oils.*

Figures 7(a) and (b) show a bar chart of the predicted outputs, $\mathbf{Y}_{\text{prediction}}$, for the chemical discrimination tasks for a group of aliphatic alcohols and a group of essential oil respectively. Distinct discriminations among testing samples have been achieved with less than ±10% error standard deviation as is shown by the error bars in Fig. 7. Using ±10% as a thresholding condition, a 100% classification accuracy was achieved within the group of alcohols and a classification accuracy of 94% was achieved within the group of essential oils. The confusion matrices for these classifications are shown in Figs. 8(a) and (b) for the group of aliphatic alcohols and the groups of essential oils respectively. No testing samples were misclassified by the optimised neuromorphic sensing system, and only a single sample was unclassified due to a $\mathbf{Y}_{\text{prediction}}$ value outside the thresholding limits.

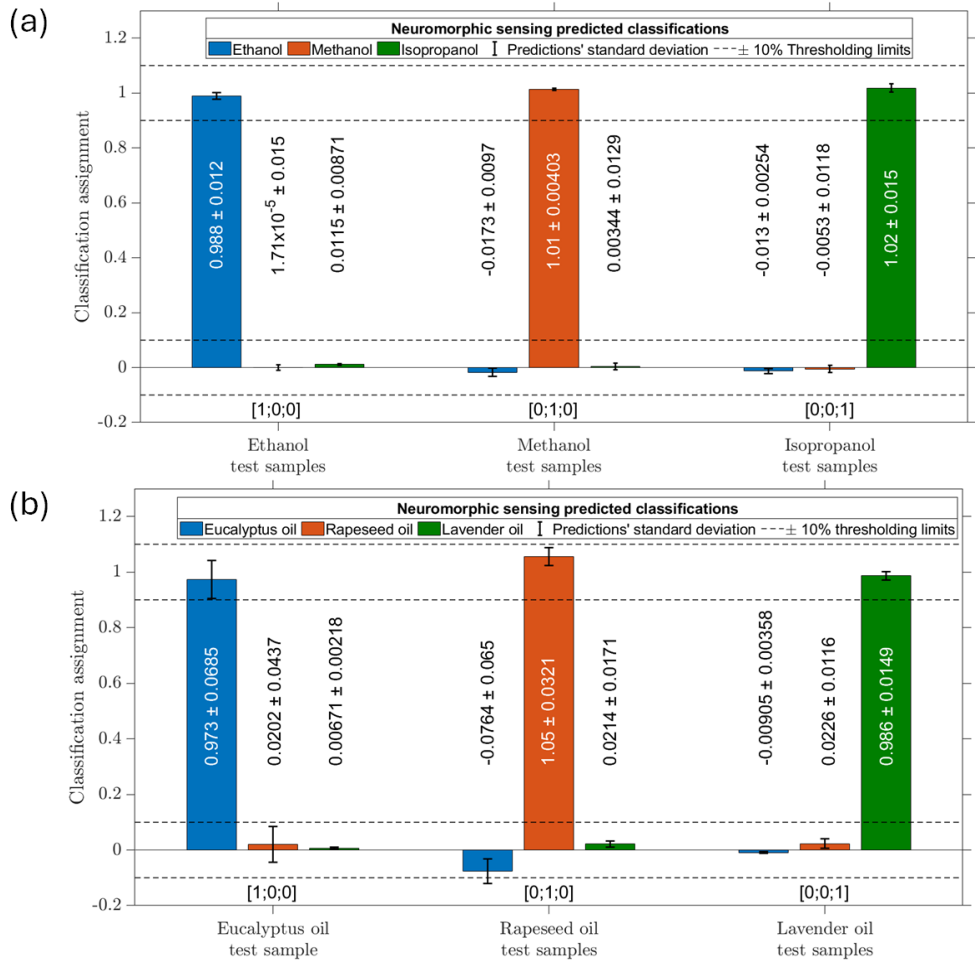

*Figures 7(a) and 7(b) depict the* discrimination *results for groups of aliphatic alcohols and essential oils, respectively. Within this classification ± 10% thresholding limits were applied in order to judge the system classification quality.*

*Figure 8. Confusion matrices of the discrimination results for neuromorphic sensing of (a) the group of aliphatic alcohols and (b) the group of essential oils.*

Here, we also report the observation that the neuromorphic sensing system developed here was influenced by ambient changes of the environment, including noise (electrical noise, thermal noise, acoustic and mechanical vibrations) and thermal component drifts (of the laser and MZM). To exemplify this, Fig. 9 presents a histogram of the *NMSE* values for 100 independent instances of classification for the group of aliphatic alcohols, carried out using a fixed set of optimised EORC parameters, *P* and $\gamma$ (see Fig. 5(a)), at randomly selected times over the duration of a week, with the same training and testing datasets used throughout. The *NMSE* mean for an optimised neuromorphic sensing system applied to classification of the group of aliphatic alcohols was 0.0148 and had an error standard deviation $\sigma = 8.73 \times 10^{-4}$. We believe that the classification robustness of the neuromorphic sensing system could be significantly improved by using low-noise photodetectors and RF-amplifiers, performing signal averaging to enhance the signal-to-noise ratio of the signal at the readout layer, $x(t)$, and by employing a more stable (less jittering) control system unit for the VOA and RF-amplifier.

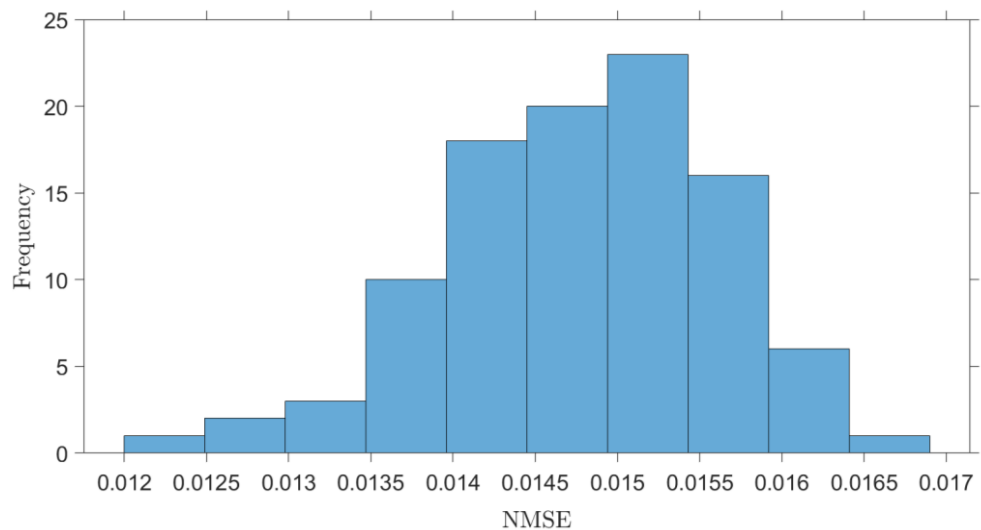

*Figure 9: A histogram depicting the NMSE for 100 runs of aliphatic alcohol discrimination with the constant optimised parameters.*

Furthermore, we also investigated the impact of the number of masks, $N_x$, used to multiplex the input signal, $u(t)$, on the system performance, *NMSE*. Figure 10 shows the *NMSE* for the discrimination of the group of aliphatic alcohols for various numbers of masks, $N_x$, and the error bars depict the standard deviation from the mean values. It confirms that a higher number of masks, $N_x$, improved the accuracy of the neuromorphic sensing system as suggested by previous publications [1,25,26]. The available number of masks was limited by equipment employed in the experiment and a steady increase in the performance of the neuromorphic sensing system is observed until $N_x = 50$. Better performance may be possible, however the standard deviation of the *NMSE* increases after $N_x = 50$, likely due to the ability of the AWG to resolve individual masks becoming increasingly compromised. The rate of increase of system performance as a function of $N_x$ until $N_x = 50$ was observed to be in agreement with the numerical analysis of the same system in previously published simulation work [1].

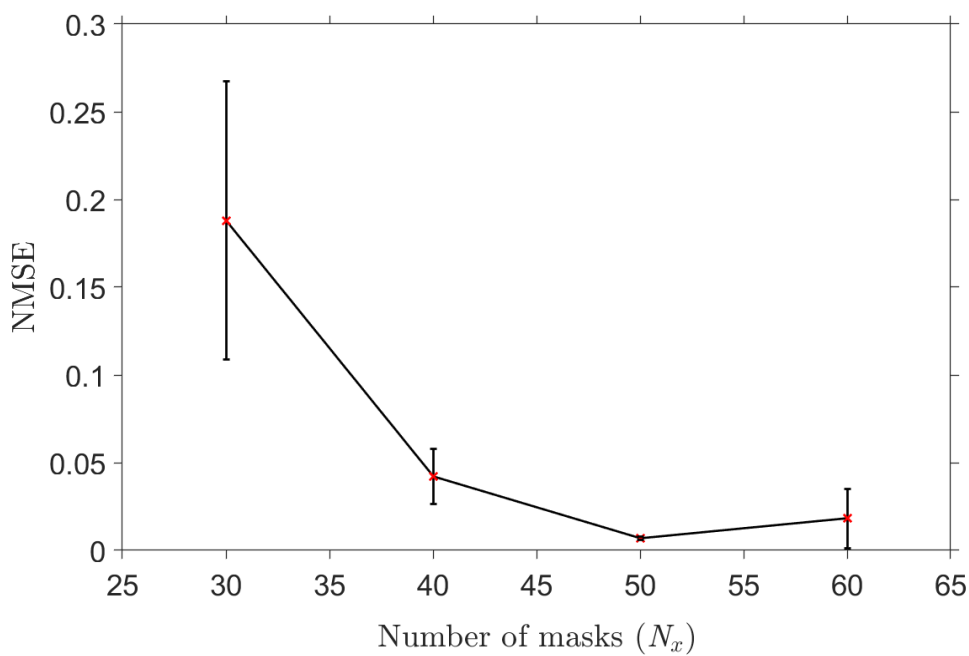

*Figure 10. The NMSE for the classification of the group of aliphatic alcohols as a function of the number of masks, $N_x$. The error bars denote the standard deviation of the NMSE over 10 runs from the mean values and the solid line connects the mean values of NMSE for each value of $N_x$.*

## 4. Conclusions

This paper has demonstrated for the first time an experimental implementation of a neuromorphic sensing system. This system consists of four distinct parts, namely the control layer, the sensing and pre-processing layer, the EORC kernel, and the post-processing and read-out layer. A methodology has been demonstrated for the automatic optimisation of the sensing system for the chemical classification task. The performance of the system has been evaluated for a group of three aliphatic alcohols and a group of three essential oils. It has been shown that using a thresholding limit of just ± 10%, with a training set of as low as 48 essential oils, yields a classification success of 94%; a perfect classification was also achieved for aliphatic alcohols with a training set of 72 samples. The bifurcation of the system was studied and validated by the numerical simulated results. Finally, the stability of the system was studied, and a range of operational parameters suggested for optimal stability.

**Data availability**

The data gathered in the experimental work of this study, supporting the findings of this work are available from the corresponding author upon reasonable request.

**Acknowledgements**

This work was partially supported by the Engineering and Physical Sciences Research Council (EP/V048937/1).